\documentclass[conference]{IEEEtran}
\IEEEoverridecommandlockouts
\usepackage{cite}
\usepackage{amsmath,amssymb,amsfonts}
\usepackage{algorithm}
\usepackage{graphicx}
\usepackage{textcomp}
\usepackage{xcolor}
\usepackage{algpseudocode}
\usepackage{url}
\usepackage{color,soul}
\usepackage{tikz}
\usepackage{multirow}
\def\BibTeX{{\rm B\kern-.05em{\sc i\kern-.025em b}\kern-.08em
    T\kern-.1667em\lower.7ex\hbox{E}\kern-.125emX}}
\begin{document}

\title{Edge Rendering Architecture for multiuser XR Experiences and E2E Performance Assessment}

\author{\IEEEauthorblockN{Inhar Yeregui, Daniel Mej\'ias, \\Guillermo Pacho, Roberto Viola}
\IEEEauthorblockA{\textit{Fundaci\'on Vicomtech} \\
\textit{Basque Research and Technology Alliance}\\
San Sebasti\'an, Spain\\
iyeregui@vicomtech.org}

\and
\IEEEauthorblockN{Jasone Astorga}
\IEEEauthorblockA{\textit{Department of Communications Engineering} \\
\textit{University of the Basque Country}\\
Bilbao, Spain\\
jasone.astorga@ehu.eus}

\and
\IEEEauthorblockN{Mario Montagud}
\IEEEauthorblockA{
\textit{i2CAT Foundation and} \\
\textit{Universitat de València} \\
Barcelona, Spain \\
mario.montagud@i2cat.net}

}

\maketitle

\begin{abstract}
Holographic communications are gaining ground among emerging eXtended-Reality (XR) applications due to their potential to revolutionize human communication. However, these technologies are characterized by higher requirements in terms of Quality of Service (QoS), such as high transmission data rates, very low latency, and high computation capacity, challenging current achievable capabilities. 
In this context, computation offloading techniques are being investigated, where resource-intensive computational tasks, like rendering XR experiences, are shifted from user devices to a separate processor, specifically an Edge Computing instance. This paper introduces an Edge Rendering architecture for multiuser XR experiences, implements it on top of widely employed XR and Web technologies, and proposes a method based on image and audio processing to evaluate its performance in terms of end-to-end media streaming latency, inter-device, and intra-media synchronization when employing different access networks.
\end{abstract}


\section{Introduction}
Significant advances in eXtended-Reality (XR) technologies, such as augmented (AR) and virtual (VR) reality, have pushed the boundaries of mobile networks in terms of Quality of Service (QoS) requirements. These applications place significant demands on networking infrastructure, such as ultra-low latency and increased bandwidth \cite{clemm2020toward}. 6G connectivity aims to address these requirements by implementing the necessary technologies to materialize fully-fledged ubiquitous mobile ultra-broadband communications \cite{9806418}.


Rendering these XR experiences requires high computation capabilities not always available on users' devices,
and computation-expensive processing may heavily impact the battery life when employing mobile devices such as smartphones. Combined with the high QoS requirements, these factors pose limitations for achieving widespread adoption of these technologies due to challenges in interoperability and scalability.



To address these obstacles and pave the way for expansion on a larger scale of XR, a strategic solution entails harnessing the capabilities of cloud and edge computing for offloading the complex tasks, such as decoding 3D content or rendering it, to a remote server, which dynamically generates a 2D perspective from volumetric video based on the user's head movements \cite{3GPP2022}. Subsequently, the server compresses the rendered texture into a 2D video stream and transmits it across a network to a client device. The client's task is then simplified to decoding the video stream, conducting minimal rendering, and presenting it to the user. Consequently, the computational load on the client side is significantly diminished \cite{qian2019toward}. 




However, a notable drawback of remote rendering is the heightened media transmission latency, due to the delays introduced by the network and media processing. Latency frequently results in virtual objects lagging or moving unpredictably relative to their intended positions. Numerous studies indicate that increased interaction latency may diminish the perceptual stability of objects and impede task performance in AR and VR environments \cite{9112752,10089176}. Moreover, in interactive applications of real-time audio and video transmission, such as multiparty videoconferencing
or other media synchronization applications like networked music interaction
or cloud gaming,
the overall one-way delay needs to be short to give the user an impression of real-time responses and to provide natural interaction. Multi-access Edge Computing (MEC)
was created to bring cloud resources near mobile users and substantially diminish latency. Consequently, deploying a rendering server on Edge as a Virtual Network Function (VNF) presents a distinct opportunity to fulfill the demands of highly sensitive to latency applications, such as XR conferencing \cite{9951054}.

Another relevant aspect in these kinds of scenarios is synchronization  \cite{montagud2018mediasync}. The user devices need to be synchronized so that they experience the actions or events at the same time.  Inter-device synchronization is essential for creating a shared and cohesive experience in collaborative applications. Additionally, intra-media synchronization plays a crucial role. According to \cite{steinmetz1993human}, for lip-sync asynchrony, which represents the asynchrony between video and audio, levels below 80 ms are unnoticeable, but levels above 160 ms become unacceptable. This intensifies the need for solutions that ensure reduced end-to-end latency and mechanisms that enable inter-device and intra-media synchronization assessments.

This paper introduces an edge rendering architecture for XR experiences where users can interact with a 3D scene and with each other, and it identifies QoS limitations by evaluating its performance in terms of end-to-end latency and synchronization in three different access network scenarios. In particular, this paper focuses on measuring the video and audio latency to assess the inter-device asynchrony and the intra-media asynchrony, also known as lip-sync, of each user, employing Ethernet, Wi-Fi, and 5G-SA access networks. The paper contributes to the state of the art in the following ways:

\begin{itemize}


    \item Design of a modular Edge Rendering architecture for multiuser XR experiences where users can interact with the 3D environment and each other by exchanging real-time video, audio, and 6 degrees of freedom (6DoF) data through an XR WebPlayer. 

    \item A method based on image and audio processing to analyze the QoS of multiuser XR experiences delivered through Web Real-Time Communication (WebRTC). This method enables evaluating the E2E video and audio latency and inter-device and intra-media synchronization.

    \item Assessment of E2E metrics when considering different access networks such as Ethernet, Wi-Fi, and 5G-SA on top of the implemented architecture and methodology. From this analysis, we will identify any limitations and possible improvements. 

    
\end{itemize}

The rest of the paper is structured as follows. Section \ref{sec:RelatedWork} reviews related work in the domain of remote rendering solutions applied to XR applications. In Section \ref{sec:architecture}, we elaborate on our modular architecture for the XR edge rendering solution. Section \ref{sec:qosmethod} describes the proposed methodology to measure the QoS of the solution. Section \ref{sec:evaluation} focuses on describing the details of the experimental assessment. Finally, some conclusions and future work are presented in Section \ref{sec:conclusions}.

\section{Related Work}
\label{sec:RelatedWork}


This section explores current approaches in architecture design, module implementation, and performance assessment, evaluating their strengths and weaknesses.


Authors in \cite{bassbouss2023metaverse} or \cite{gul2020low} propose different architectures, all implementing remote rendering, but each focuses and enhances specific aspects, such as device homogeneity or user interaction. In \cite{huang2023scaxr}, the authors also focus on remote rendering techniques for their scalability-related experiments. This approach enhances device compatibility, making it accessible on any device capable of running a web browser, so it can be directly received and displayed by a VR headset. We have aimed to focus on user interaction with the scene and among themselves, so we have ensured that users can share video and audio in real-time, along with 6DoF information.

Real-time communications prioritize instantaneity, making the QoS and Quality of Experience (QoE) in multiuser XR experiences heavily reliant on the end-to-end latency of audio, video, and data streams \cite{9136591}. 
Regarding the QoS evaluation of a remote rendering system, we consider that one of the key aspects to analyze is the end-to-end latency, also known as motion-to-photon (M2P) latency, in the case of the video, and mouth-to-ear (M2E) latency, in the case of the audio.  Some previous studies have focused their research on assessing a remote renderer's performance by characterizing the underlying network's delays with ping latency \cite{shi2019mobile} \cite{zhang2019analysis}, assuming that network latency is the key driver of overall latency. When a remote renderer is used, in most cases, the content is transmitted to users via WebRTC, as it provides low latency. The WebRTC Stats API provides information about these data transmissions, such as round-trip time (RTT), as described in \cite{8848499}. However, this poses a limitation when measuring the M2P and M2E latencies of communication between two users connected to a remote renderer, as in this case, there would be two WebRTC connections, i.e., two hops, and it would be necessary to manually add the latency values provided by the WebRTC Stats API for each connection. Conceptually, the most commonly used method to overcome this constraint involves inserting time references into multimedia streams and analyzing them upon arrival.

In this context, authors of  \cite{gul2020low} or \cite{7848838} employ methods in which timestamps are introduced into the video stream, and the player is capable of detecting and computing the time at which this content was created, thereby calculating the M2P latency. Concerning M2E latency, the authors of \cite{sacchetto2021jacktrip}  and \cite{doi:10.1177/1071181321651332} also incorporate temporal references in the form of beeps or pulses and employ external hardware, such as recorders, they compare the moment at which the sounds are generated with the moment at which the receiver reproduces them, thereby calculating the M2E latency. Upon reviewing the state-of-the-art methods for measuring M2P and M2E latency, it has been observed that, firstly, there is a lack of standardized methodologies, and secondly, the employed techniques often suffer from imprecision when measuring both M2P and M2E latencies simultaneously. The lack of automatic monitoring capabilities without external hardware has also been identified.  Additionally, it has been observed that, in general, methods are sought that do not introduce communication overheads caused by extra signaling or out-of-band signaling to evaluate QoS.

Following this approach, we propose a service to provide timestamps in the form of Quick Response (QR) codes and tones of specific frequencies as a method for QoS performance assessment purposes. The QR codes are easily inserted into the video stream, while the tones are inserted into the audio stream, and both are created synchronously with each other. This works in collaboration with detectors for these timestamps implemented in the player used by the users, which is designed to automatically export these metrics.

\section{XR edge rendering Architecture}
\label{sec:architecture}

\begin{figure}[t!]
\centerline{\includegraphics[width=0.5
\textwidth,keepaspectratio]{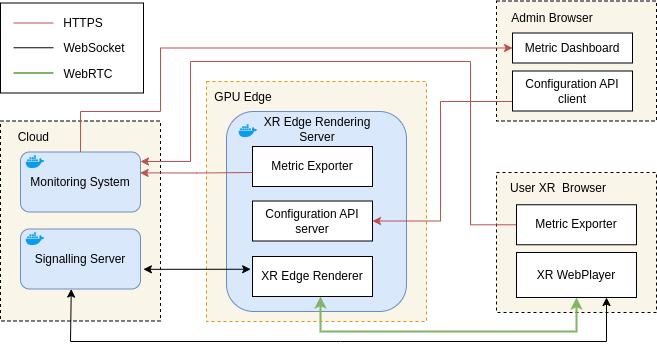}}
\caption{General architecture of the system.}
\label{fig:architecture}
\end{figure}

When designing a multiuser XR experience system, many functional requirements must be considered to ensure proper functioning. First, the client has to get the user's video, audio, and 6DoF data in real time from the user's input devices and send it to the server where the VR scene is running. Then, on the server side, advanced processing has to occur, applying the received data to generate a dynamic VR scene. This scene is then rendered for each user's eyes, ensuring a personalized and immersive experience. Post-rendering on the edge, the images must be encoded and sent back to the user devices, which decode and display them in their VR headsets. A carefully crafted modular architecture has been devised to address these requirements, as Figure \ref{fig:architecture} shows.


While edge rendering enhances the user experience by delivering high-quality immersive content at a practical bitrate and reasonable complexity, ensuring low latency is crucial for seamless user interaction. This is because the actual rendering for a specific viewpoint occurs not at the client but at the edge device. The XR Edge Renderer is responsible for carrying out this task. As seen in Figure  \ref{fig:architecture}, it is located at the XR Edge Rendering Server, which runs on the GPU Edge. 
Consequently, both the user input content (video and audio) and the content rendered in the edge must be transmitted in real-time using a low-latency protocol such as the Real-time Transport Protocol (RTP), which is suitable for real-time communications (RTC). The system presented in this paper is implemented based on WebRTC, which enables achieving ultra-low latency by utilizing a peer-to-peer connection between the client and server. Moreover, WebRTC is extensively embraced by various web browsers, enabling our system to seamlessly cater to multiple platforms. To facilitate a WebRTC communication, the system architecture must include a Signalling Server. This server plays a pivotal role in facilitating communication by overseeing the negotiation process known as the Interactive Connectivity Establishment (ICE). ICE negotiation is a crucial step in establishing a connection between two WebRTC peers, involving exchanging network information to determine the optimal path for data transmission.
 
 The architecture has also been provided with a Monitoring System, where both server-side and client-side metrics are collected by employing specific Metric Exporters. Including performance metric collection further enhances the system's capabilities by providing valuable analytics for continuous improvement and optimization. Moreover, a configuration application programming interface (API) has been deployed, enabling an administrator to make decisions based on the monitored metrics, such as adapting the bitrate, framerate, or resolution of the video streams sent to the users.
 
 This architectural design satisfies the specified requisites and ensures a robust and efficient holographic conferencing platform.

\section{QoS for multiuser XR experiences}
\label{sec:qosmethod}

Accurate latency measurement between multimedia content transmission modules requires synchronization through Network Time Protocol (NTP). A stand-alone NTP has been implemented to align the clocks of all system modules, ensuring a common time reference for precise latency assessment \cite{8638081}. This synchronization is essential to eliminate time discrepancies among modules, allowing for reliable timestamping and coordinated actions. 

To assess the performance of the XR Edge Renderer in terms of end-to-end latency and synchronization, we developed a method based on image and audio processing in the XR WebPlayer.
We developed a web-based application that generates and displays a QR code based on the current timestamp every 10ms. Concurrently, it emits audio pulses of specific frequencies synchronized with the current time. The machine's clock, synchronized via NTP, serves as the temporal reference. 

Let's consider a scenario where two clients engage in a multiuser XR experience utilizing the XR WebPlayer. In this scenario, one of the clients, referred to as the presenter, not only shares their media content but also consumes it. The other client, referred to as the viewer, consumes the shared content. The clients' uplink channels are used to inject customized data (video or audio). In the case of video, the QR code is inserted in the uplink video stream of the presenter and displayed in the virtual scene. When the viewer displays the video frame with the QR code, it will detect and extract the embedded timestamp while obtaining the current timestamp from its local clock. This has been implemented using the Insertable Streams API \cite{blum2021webrtc}, which gives us access to the data within the MediaStreamTrack and allows us to pass it through a custom function that we have implemented. 
In this case, we use the function to extract and decode the timestamp information. This way, we have the timestamp of the moment the presenter captured a video frame and the moment the viewer viewed it on their headset. The difference between these two timestamps represents the entire system's M2P or end-to-end frame/video latency. 

A similar method has been followed to measure the audio latency. The presenter injects the synchronized audio tones into its uplink audio channel. Then, this audio is received by the viewer and processed to detect these samples, identifying each of them by their frequency. This has been implemented using the Web Audio API \cite{smus2013web}. The used approach to calculating the frequency of a signal involves accessing the raw sound data directly and employing autocorrelation. This technique entails comparing segments within the signal, such as a sound wave, against delayed replicas of themselves. By systematically adjusting the delay offset and analyzing the similarity between the signal and its delayed versions, we can determine the offset at which the signal pattern approximately repeats. This offset corresponds to the period of the sound wave, enabling straightforward derivation of its frequency. 
The timestamp of the moment when the audio samples are reproduced is observed and compared with the timestamp of the moment when they have been created, which is known in advance, depending on their frequency. The difference between the two timestamps represents the E2M or end-to-end audio latency.

 \begin{figure}
     \centering
     \includegraphics[width=1\linewidth]{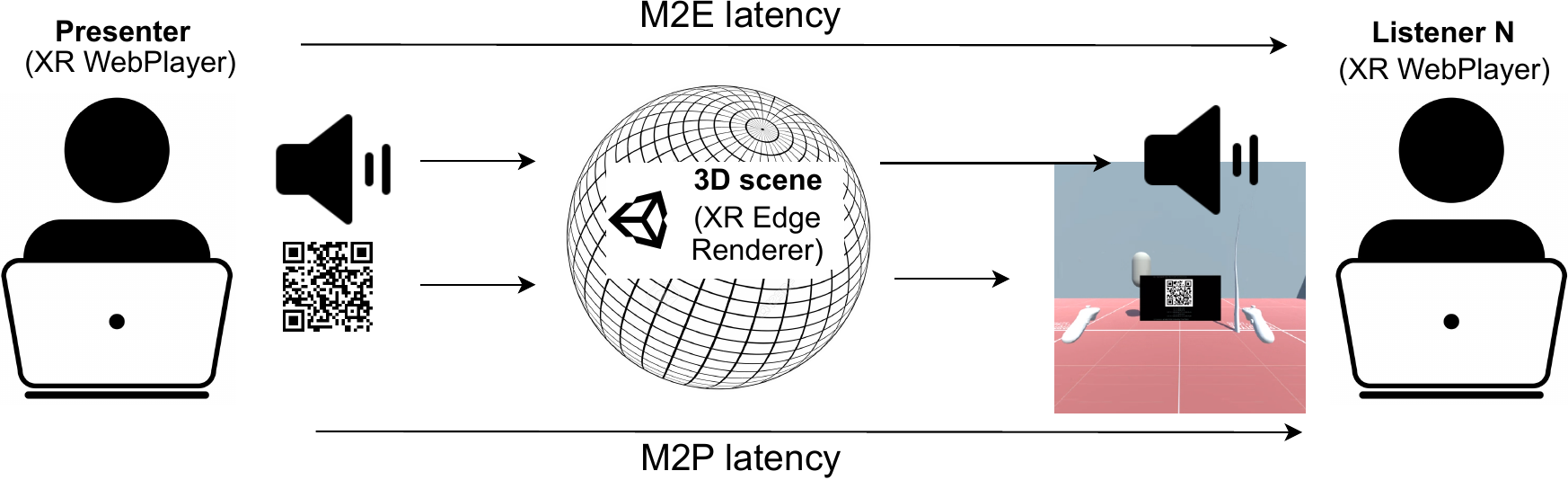}
     \caption{Testbed setup.}
     \label{fig:testbed}
 \end{figure}

The XR WebPlayer then exports all these metrics so that the monitoring system can collect them. Monitoring these latencies is valuable for observing the delay difference that audio and video streams might suffer and helps to assess the synchronization between video and audio in a single player. Additionally, it proves beneficial for assessing the synchronization among different players in scenarios involving multiple participants. These metrics can be employed for decision-making purposes to uphold or enhance the QoS and QoE of the service. For instance, they can be leveraged through the configuration API to adjust parameters such as bitrate, framerate, or resolution of the videos sent to the users.

\section{Experimental assessment}
\label{sec:evaluation}

    
\subsection{Testbed setup}
\label{sec:testbed}

The remote rendering service has been implemented utilizing the Unity Render Streaming plugin, which can be integrated into Unity Projects. A virtual 3D environment has been created using the Unity game engine, and it is responsible for bundling all the user's XR experiences. It oversees the management of each connected client through a standardized signaling process. This process includes the transmission of rendered streams, reception of audio and video streams, and the handling of input events. It has been containerized and runs headless in a GPU-enabled Edge Server. Vulkan Graphics API has been implemented to enable full GPU acceleration in the container. 

The heterogeneity of the users is one of the requirements we have been seeking during the implementation phase. This has led us to choose native web technologies for the XR WebPlayer. To enable interactive multiuser XR experiences, the web player needs to be able to capture users' video, audio, and 6DoF interaction data and send it to the XR server so that it can be integrated into the 3D virtual scene. Moreover, it must display personalized 2D videos received from the server and in VR mode. This has been implemented with A-Frame \cite{aframe}. 
The VR scene is crafted by leveraging two 2D videos as a foundational element.

The tests were performed by sequentially connecting 5 players to the system, with approximately one player added every minute. Laptop computers with WebXR browsers were utilized to run the players due to the absence of VR headsets, as we required 5 devices operating simultaneously. Initially, the first player was the presenter, responsible for transmitting multimedia video and audio content to the virtual scene. The presenter operated alone in the session throughout the first minute, as shown in Figure \ref{fig:testbed}. Then, additional users/clients were added every minute until reaching a total of 5 users, who assumed the role of viewers. Each test lasted 300 seconds and was repeated 3 times, each employing a different access technology. In the first scenario, all clients connect to the remote server via Ethernet. We use the results obtained in this test as the ground truth to aim for, as it represents the best-case scenario after configuring the encoder, queues, data processing, etc., to achieve minimum latency. In the second case, the clients connect to the server via Wi-Fi; in the third case, we use a 5G Stand Alone network.


\subsection{Results and discussion}
\label{sec:results}

\begin{figure}[!t]
  \centering
  \begin{minipage}[b]{0.5\textwidth}
    \includegraphics[width=\textwidth]{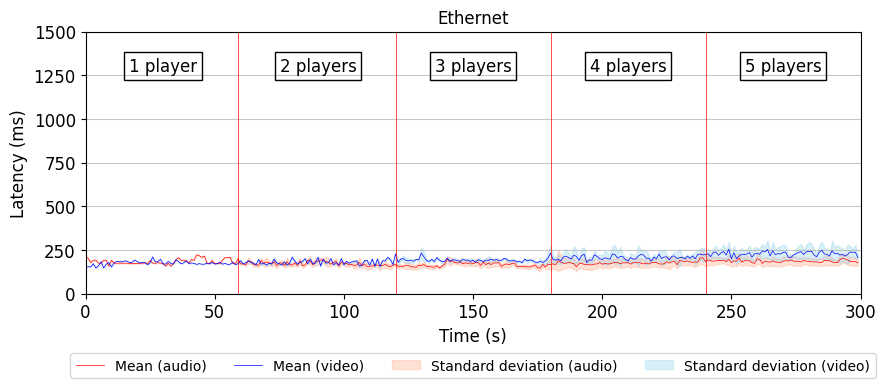}
  \end{minipage}
  \hfill
  \begin{minipage}[b]{0.5\textwidth}
    \includegraphics[width=\textwidth]{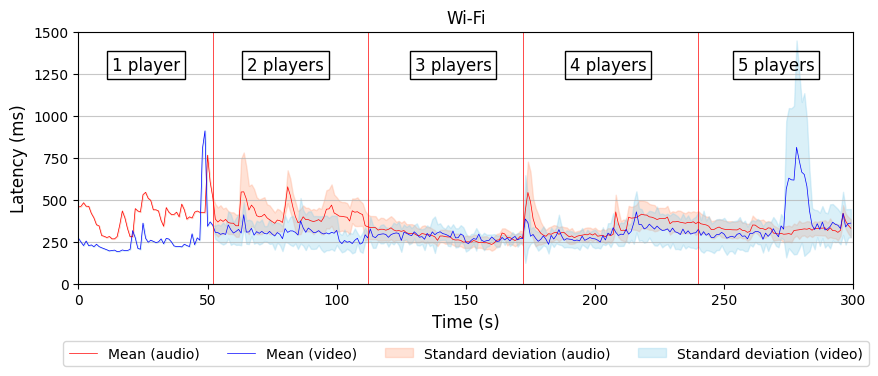}
  \end{minipage}
   \hfill
  \begin{minipage}[b]{0.5\textwidth}
    \includegraphics[width=\textwidth]{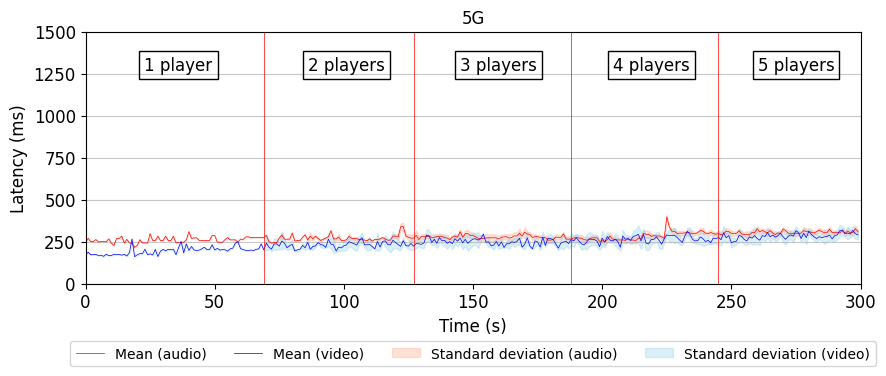}
  \end{minipage}
  \caption{M2P and M2E latency.}
  \label{fig:e2e-lat}
\end{figure} 

\begin{figure*}[t]
  \centering
  \begin{minipage}[b]{0.325\textwidth}
    \includegraphics[width=\textwidth]{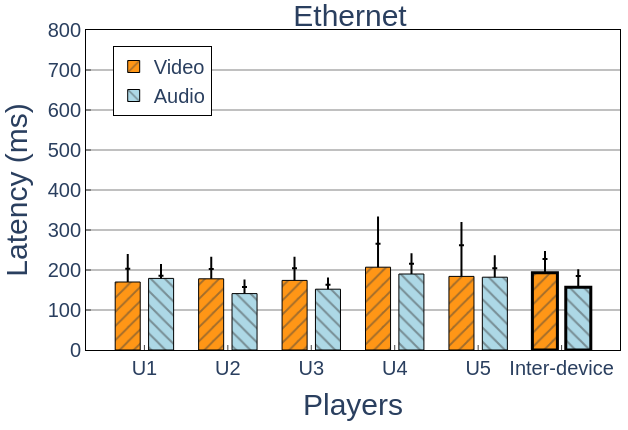}
  \end{minipage}
  \hfill
  \begin{minipage}[b]{0.325\textwidth}
    \includegraphics[width=\textwidth]{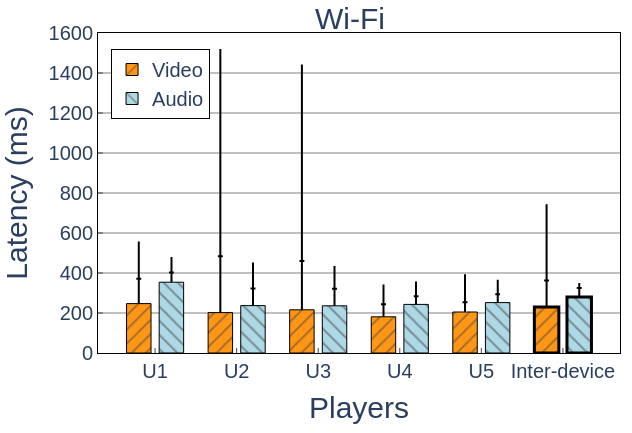}
  \end{minipage}
   \hfill
  \begin{minipage}[b]{0.325\textwidth}
    \includegraphics[width=\textwidth]{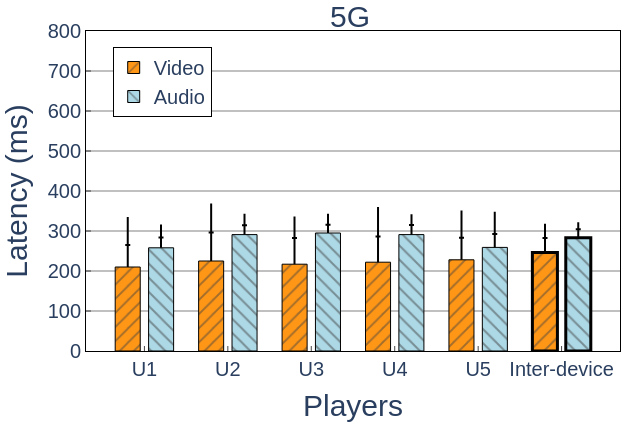}
  \end{minipage}
  \caption{Inter-device asynchrony.}
  \label{fig:inter-device}
\end{figure*} 

\begin{figure*}[t]
  \centering
  \begin{minipage}[b]{0.3\textwidth}
    \includegraphics[width=\textwidth]{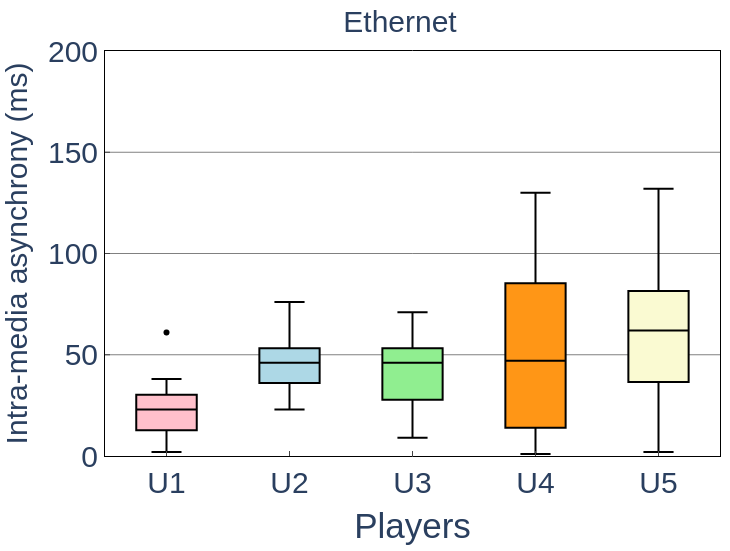}
  \end{minipage}
  \hfill
  \begin{minipage}[b]{0.3\textwidth}
    \includegraphics[width=\textwidth]{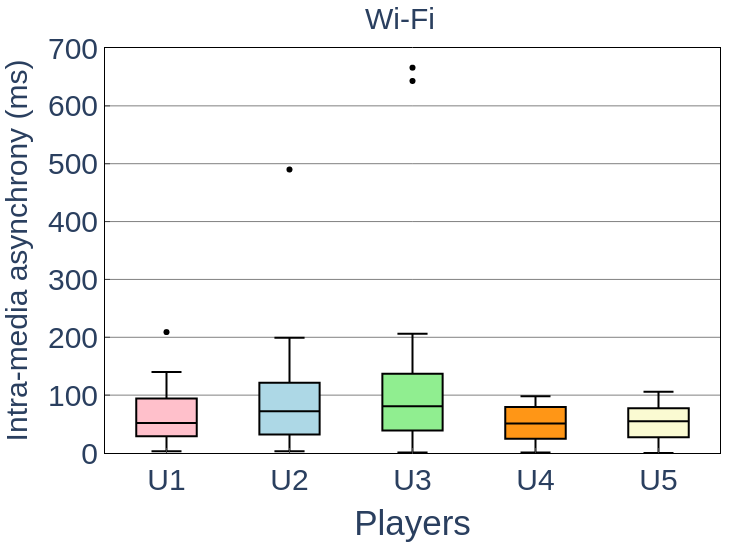}
  \end{minipage}
   \hfill
  \begin{minipage}[b]{0.3\textwidth}
    \includegraphics[width=\textwidth]{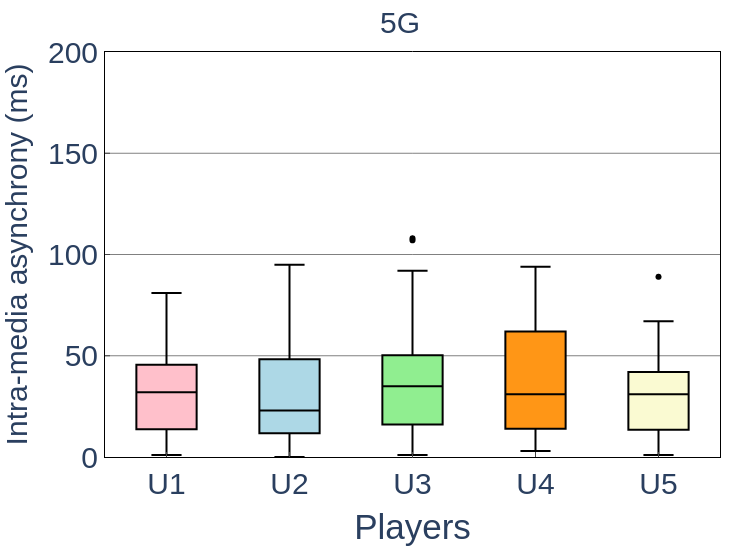}
  \end{minipage}
  \caption{Intra-media asynchrony.}
  \label{fig:intra-media}
\end{figure*} 

This section analyzes the outcomes obtained from the testing of the implemented solution.  Measurements concerning M2P and M2E latency are shown in Figure \ref{fig:e2e-lat}. These show the connected clients' mean value and standard deviation of both video and audio latency in each time slot. The mean values are shown as a solid line, while a shadow around the mean value represents the standard deviation. The mean value provides information about the system's performance limitations regarding media streaming latency when implementing different access network technologies and when various clients simultaneously participate in an XR session. Furthermore, the standard deviation describes the variability of latency experienced by different users. The greater the standard deviation, the greater the variability, indicating more asynchrony among them.

In the case of Ethernet, the first graph of Figure \ref{fig:e2e-lat} shows that both average video and audio latencies remain under 250 ms throughout the entire test. The mean latency values in both streams show no significant impact from the increase in the number of concurrently connected users. However, the standard deviation rises, particularly in the final slot representing 5 connected users, indicating increased video and audio latency variability.

Regarding the experiment conducted with users connected to the network via Wi-Fi, shown in the second graph, a discernible degradation is observed in both video and audio latencies and the variability or asynchrony among different users. In this context, upon examining the latency trends over time and considering the number of users connected at each instance, it can be inferred that the degradation of service quality is primarily influenced by temporal variability rather than the number of users. Notably, the instance with two users exhibits higher latency and asynchrony than scenarios with three or four connected users.

In the case of the 5G test, as depicted in the third graph, we observe values similar to those obtained in the Ethernet test. However, in this scenario, latencies are slightly elevated, as anticipated, owing to inherent limitations of wireless technology when compared to wired counterparts. Notably, regarding asynchrony, the number of connected users does not exhibit a significant impact.


From these three graphs, it seems that the number of players, up to a maximum of 5 connected simultaneously, does not negatively impact the system's performance in terms of E2E latency and, therefore, the asynchrony between users. Consequently, the remaining analysis focuses on the last slot of the test, when there were 5 users connected simultaneously, as we believe this to be the most interesting scenario to examine inter-device and intra-media asynchrony.

Figure \ref{fig:inter-device} summarizes each user's audio and video latencies when 5 users were connected. The bars represent the minimum latency recorded for each user during the test, while the upper whiskers illustrate the variability, or jitter, of these data, representing the highest value and the average of all values. Then, a composite of all player samples is depicted, providing insights into inter-device asynchrony for both video and audio. The bar signifies the mean value of the minimum latencies recorded at each test interval, while the whisker delineates the range of asynchrony, showcasing both the maximum and mean values. 

If we look at the graphs in Figure \ref{fig:inter-device}, we can observe that the performance of Ethernet and 5G technologies is quite similar concerning inter-device asynchrony. Although 5G shows slightly more latency, the difference in latencies obtained in both cases is very small. In contrast, the system performs poorly when users use the access network via Wi-Fi. In Table \ref{tab:inter-device-table}, we can see more details about the values obtained. On the one hand, we see that the lowest average values for both video and audio are obtained in the case of Ethernet, followed by 5G, and ending with Wi-Fi. Regarding inter-device asynchrony, in the case of Ethernet or 5G, we obtain values below 75 ms. To reach this value, first, the asynchrony of each user device is calculated at each moment, using the one with the minimum latency as a reference. To do this, the latency difference obtained by each user device relative to the minimum is calculated. Then, the average of these values is taken, obtaining the inter-device asynchrony at each moment. The maximum of these values represents the inter-device asynchrony. In the case of Wi-Fi, we see that the inter-device asynchrony for video rises to 514.2 ms. The values obtained in the Ethernet and 5G scenarios are under the time corresponding to 2 video frames, achieving the synchronization target \cite{montagud2018mediasync}.

\begin{table}[t]
\caption{inter-device asynchrony summary.}
\centering
\def\arraystretch{1.2}
\label{tab:inter-device-table}

\begin{tabular}{lc|c|c|c|}
\cline{3-5}
                                                                  &       & \textbf{Ethernet} & \textbf{Wi-Fi}   & \textbf{5G}     \\ \hline
\multicolumn{1}{|l|}{\multirow{2}{*}{\textit{Average E2E Latency (ms)}}}   & Video & 227.54   & 362.46 & 282.67 \\ \cline{2-5} 
\multicolumn{1}{|l|}{}                                            & Audio & 185.22   & 324.59 & 304.17 \\ \hline
\multicolumn{1}{|l|}{\multirow{2}{*}{\textit{Inter-device asynchrony (ms)}}} & Video & 54.2     & 514.2  & 71.8   \\ \cline{2-5} 
\multicolumn{1}{|l|}{}                                            & Audio & 45       & 69.2   & 38.8   \\ \hline
\end{tabular}
\end{table}

Another interesting aspect to analyze is intra-media synchronization, also known as lip-sync, which represents the synchronization between the video and audio of the same player. In this case, the same time frame has also been evaluated: the moment when 5 users are connected simultaneously. Figure \ref{fig:intra-media} displays the data obtained in the test in the form of box plots. They represent the central tendency, variability, and distribution of intra-media asynchrony values over time during the experiment. 

In this context, the y-axis is not the same for all three graphs, as the results obtained with Wi-Fi show outliers far from the values obtained with the rest of the technologies. Specifically, the presence of outliers in the Wi-Fi data, distant from the corresponding values observed with alternative technologies, highlights anomalous behavior within the Wi-Fi network, as asynchrony values of up to 650 ms have been recorded. Upon closer examination of Figure \ref{fig:e2e-lat}, it becomes apparent that this intra-media asynchrony can be attributed to instances where video playback experienced interruptions, leading to temporal disparities between video and audio streams. Despite these interruptions, the audio component remained largely unaffected. In the case of Ethernet, it is observed that the values of intra-media asynchrony generally remain below 100 ms, and even dip below 50 ms in the case of 5G. According to \cite{steinmetz1993human}, this confirms that intra-media asynchrony remains imperceptible to humans, for these two cases.


\section{Conclusions and Future Work}
\label{sec:conclusions}
The present work has proposed an edge rendering architecture for XR experiences implemented and validated through the method that evaluates its QoS regarding E2E latency and inter-device and intra-media asynchrony. The method for measuring and monitoring intra-media and inter-device asynchrony offers valuable insights for evaluating multiuser XR experiences. It provides a detailed understanding of service quality by assessing end-to-end latency and synchronization between audiovisual media. This approach identifies potential issues and facilitates optimization across different access networks, ultimately enhancing user experience. Real-time monitoring allows for informed decision-making and continual improvement, contributing to the advancement of multiuser XR applications. It also highlights the poor synchrony of the Wi-Fi setup when compared to 5G-SA. The future work will stress the dynamics coming from heterogeneous networks and throughputs impacting the asynchrony experienced by participants.
\section*{Acknowledgment}
This research was supported by Smart Networks and Services Joint Undertaking under the European Union’s Horizon Europe Research and Innovation programme, under Grant Agreement 101096838 for 6G-XR project, and the Spanish Ministry of Economic Affairs and This work has been funded by the Spanish Ministry of Economic Affairs and Digital Transformation and the European Union – NextGeneration EU, in the framework of the Recovery Plan, Transformation and Resilience (PRTR) (Call UNICO I+D 5G 2021, Ref. TSI-063000-2021-4 – 6G-Openverso-Holo), project). The work of Mario Montagud has been funded by MCIN/AEI/10.13039/501100011033 under Grant RYC2020-030679-I and by “the European Social Fund (ESF) Investing in Your Future”.


\bibliographystyle{IEEEtran}
\bibliography{2-main.bib}

\end{document}